\documentstyle[aps,prl,twocolumn,psfig]{revtex}

\draft
\begin{document}
\newcommand {\be}{\begin{equation}}
\newcommand {\ee}{\end{equation}}
\newcommand {\bea}{\begin{eqnarray}}
\newcommand {\eea}{\end{eqnarray}}
\newcommand {\nn}{\nonumber}

\twocolumn[\hsize\textwidth\columnwidth\hsize\csname@twocolumnfalse%
\endcsname

\title{Quantum spins and quasiperiodicity : a real space
renormalization group approach}

\author{ A. Jagannathan}
\address{Laboratoire de Physique des Solides, Universit\'e Paris-Sud,
91405 Orsay, France \\
}

\date{\today}
\maketitle

\begin{abstract}
We study the
antiferromagnetic spin-1/2 Heisenberg model on a two-dimensional bipartite
quasiperiodic structure, the octagonal tiling -- the
aperiodic equivalent of the square lattice for periodic systems. 
 An approximate block spin renormalization
scheme is described for this problem. The ground state energy and local
staggered magnetizations for this system are calculated, and 
compared with the results of a recent Quantum Monte Carlo 
calculation for the tiling. It is conjectured that the ground state energy
is exactly equal to that of the quantum antiferromagnet on the square lattice.

\end{abstract}
]

In this  paper a renormalization group transformation is used
to study the ground state of Heisenberg spins with antiferromagnetic
couplings on a two-dimensional quasiperiodic tiling. This system
poses a novel theoretical
problem, namely, the nature of quantum fluctuations in a 
structure possessing a number of exact symmetries but no translational invariance.
 While periodic systems and
disordered variants thereof have received
much attention, little is known about aperiodic quantum models in two or more
dimensions. In particular, the
real space magnetic ordering of local moments in systems
with quasiperiodic long range order remains to be elucidated, and should
present novel and complex features, different from properties of
crystalline or disordered systems. The archetypal nonfrustrated two-dimensional 
antiferromagnetic system is that of spins on the square lattice, an old
and until recently controversial problem, while the problem we consider now ,
with its
fundamentally different symmetry properties, aims to understand a
new class of unfrustrated systems.

Experimental work providing motivation for
the study of such systems comes from neutron scattering studies of the magnetic
phase in a Zn-Mg-Ho quasicrystal \cite{sato}. The
magnetic diffuse scattering of the low temperature phase
shows an icosahedral symmetry, reflecting the underlying quasiperiodicity of
this compound. 
The nature of the ground state in such a quasicrystalline medium
was recently discussed in \cite{wess} where Quantum Monte Carlo (QMC)
calculations were carried out for an antiferromagnetic 
Heisenberg model on one of the simplest two-dimensional
quasiperiodic tilings available, the octagonal tiling.
 This tiling has been frequently used for
numerical investigations of the effects of quasiperiodic modulations
in two dimensions. More detailed, analytic and numerical results are available
for one dimensional quasiperiodic models, where
  quantum spins have been considered using real space renormalization
transformation \cite{jherm}, and 
using density matrix renormalization or
by using mappings to fermionic models (see \cite{hida} and
references therein).  However, the techniques used
 are particular to one dimension and not readily generalisable
to the two-dimensional structure considered here.

 The model considered in \cite{wess} has a  Hamiltonian 
$ H = J \sum_{\langle i,j\rangle} \vec{S}_i . \vec{S}_j$
where the spins are located on vertices of the octagonal tiling, and J is
coupling along each edge.  
Spin-spin correlations
in the ground state were computed,
and the staggered local
 moment at a given site was found to depend
on the number of nearest neighbors $z$. 
We recall that
the octagonal tiling 
 has six $z$-values ranging from 3 to 8 (see ref.\cite{wess} for a picture).  
Fig.1 shows the six types of nearest neighbor configurations that occur, along
with the nomenclature used in this paper (see below).

\begin{figure}[h]
\centerline{\psfig{figure=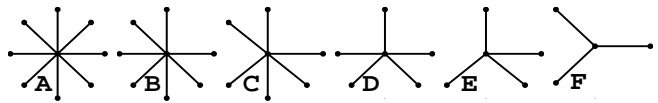,width=9cm,angle=0}}
\vspace{.3cm}
\caption{The six nearest neighbor environments on the octagonal tiling}
\end{figure}
Within each of the six families 
the local order parameters have a fine structure reflecting the differences
in the next-nearest neighbor shells, and there are
further splittings due to even longer range spin-spin interactions.
The observed $z$ dependence of the local order parameters was qualitatively
explained in \cite{wess}
by an isolated cluster or Heisenberg star approximation. 
 Doing better
requires taking into account successive shells
of next-nearest neighbors. The best way to do this
is by a renormalization group calculation, using the self-similarity
of the quasiperiodic structure. This is the aim of the present calculation.

The renormalization approach is a natural one for self-similar quasiperiodic
tilings invariant under a scale transformation or inflation
 such as the one described
by Gardner \cite{gard}
for the Penrose tiling. For the octagonal (Ammann-Beenkker)
 tiling \cite{soco},
 one can start
with tiles of some given edge length,
and reconnect a certain subset of vertices (inflation).
The redrawn tiling is then similar to
the original one, except for an overall scale factor,
equal to the golden
mean $\tau = (\sqrt{5}+1)/2$ for the Fibonacci chain,  2d Penrose tiling and
its 3d generalization, or
the silver mean $\lambda = (1+\sqrt{2})$
 in the case of the octagonal tiling. 
This structural property of
tilings has been often exploited in order to establish recurrence
relations for parameters occurring in discrete spin models, electron
hopping models, etc, as mentioned before for the one-dimensional case, 
and for some two-dimensional models \cite{godreche,moss2}, where
analytical methods remain hard to implement.

Our renormalization group is a generalization of the calculation of Sierra
and Martin-Delgado
for the square lattice \cite{sierra}, where the authors considered
blocks composed of five-spin star-shaped clusters.
On the quasiperiodic tiling, the choice of block spins is suggested by
site behavior under an inflation operation. 
Inflation results in the disappearance of low $z$ sites. After inflation the
sites that remain are those of high coordination number,
A,B,C,and $D_1$ sites, having $z=8,7,6,5$ respectively. We 
 refer to them
collectively as $\alpha$ sites. Sites that disappear have
$z=
5,4$ and 3 respectively (the $D_2$,E and F sites). 
 Note that there are
 two types of five-fold site, that behave differently under inflation
\cite{note}.
The relative  number of sites of each kind, $f_i$
is preserved 
under inflation, whereas the density of sites is reduced by $\lambda^{-2}$.
After inflation, sites have new coordination numbers $z'$ 
as indicated in the following
list of transformations :

\begin{tabular}{lcccccccccccr}
A& $\rightarrow$&A&or&B&or&C&or $D_1$&;&&&& \\
B&$\rightarrow$&$D_2$&;&&C&$\rightarrow$&E&;&&$D_1$ &$\rightarrow$& F \\
\end{tabular}

A natural choice for block spins is to consider star-clusters centered on
sites of the $\alpha$ class.
After inflation, the old $z$-blocks will become the vertices of the inflated
tiling, and new blocks defined at the high-$z$ sites, and so on. The
block spins and the couplings will renormalize to site-dependent values,
grouped according to the local
environments.
 Fig.2a shows a
central $D_1$ site, which transforms after inflation to a $z'=3$ site. The
sites remaining after inflation are shown with large dots, and the dashed
grey lines represent effective interactions between these sites. Intra-block
couplings are shown by thick lines, while inter-block couplings 
are shown by thin dotted lines.
\begin{figure}[h]
\centerline{\psfig{figure=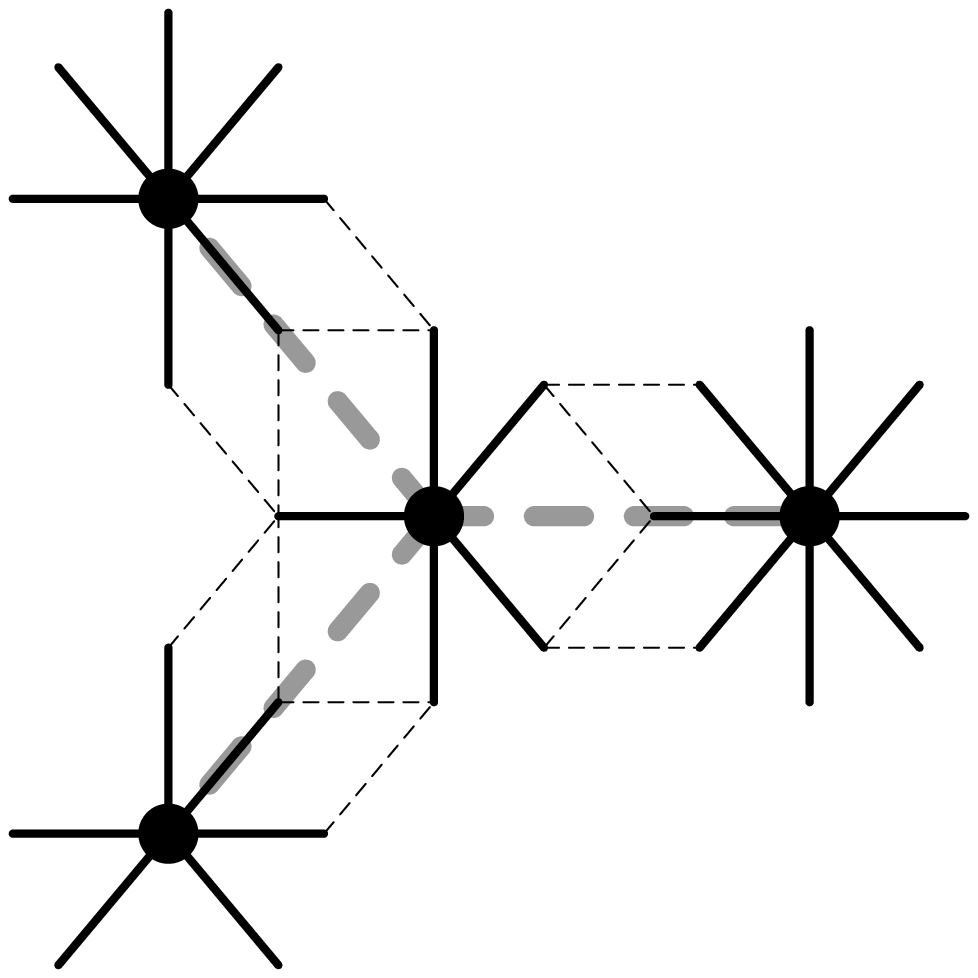,width=2.6cm,angle=0}
\psfig{figure=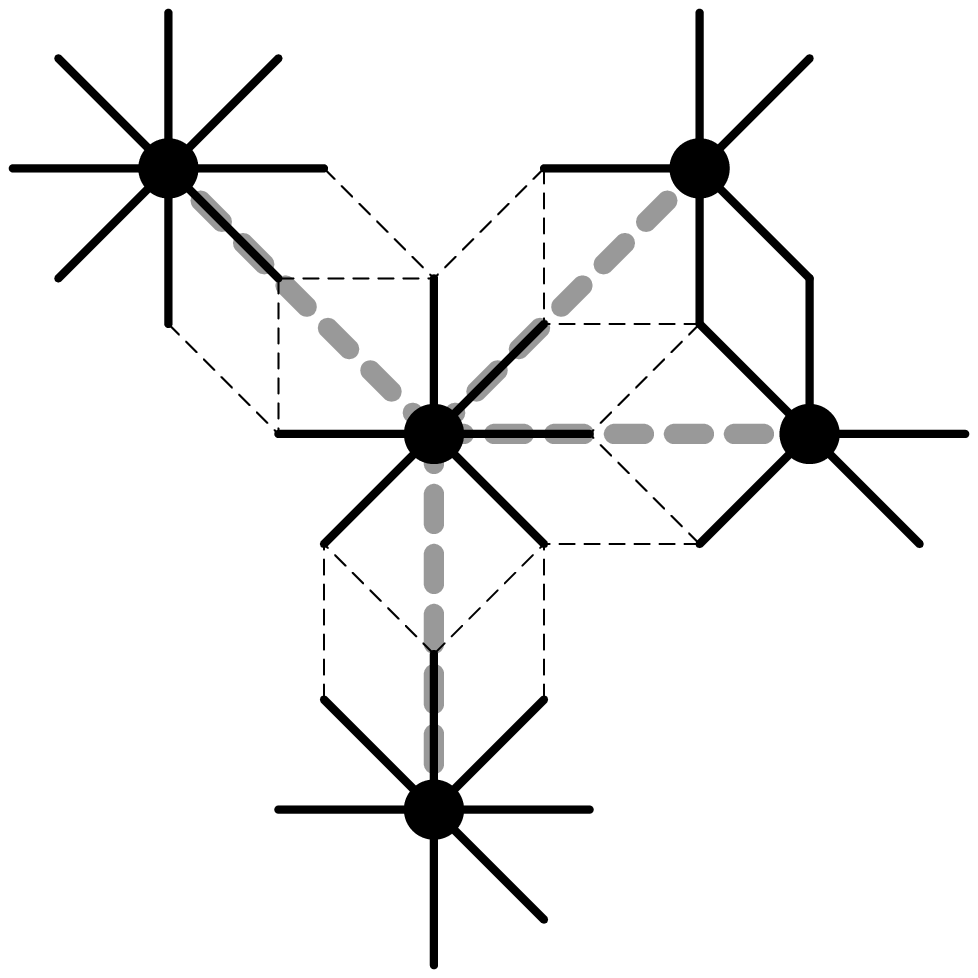,width=2.6cm,angle=0}
\psfig{figure=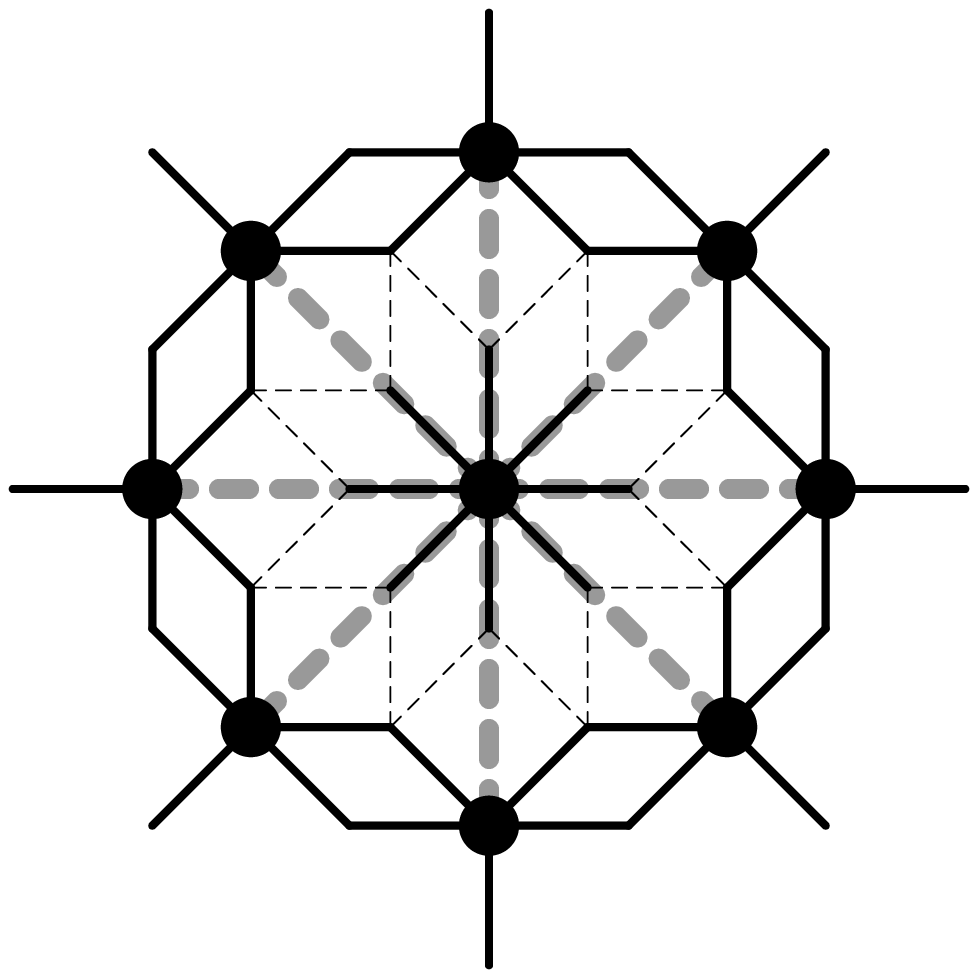,width=2.6cm,angle=0}}
\vspace{.2cm}
\caption{Block spin centers (filled circles) showing the central and all
peripheral blocks for three cases:
 (a) a $z=5, z'=3$ site
(b) a $z=6, z'=4$ site  (c) a $z= z'= 8$ site
}
\end{figure}
For an $isolated$ block spin with $z$
spins surrounding a central spin and antiferromagnetic interactions, the
cluster has a spin of $S' = (z-1)S$ in the ground state. The
energy of the isolated block can be exactly found, while for the
inter-block couplings, we will
follow the approach used in \cite{sierra} for the square
lattice (where all blocks carry the same value of $z=4$),
where one finds 
 $S'=3 S$. The
 spin
renormalization factors are taken to be equal to the classical
value $\xi^{(0)}_z =1/(z-1)$ for simplicity.
 The new block spins $S'$ are
situated on the black circles representing the sites of the
inflated lattice, while all of the nearest neighbors are decimated
in the renormalization group (RG) transformation. 

However, on the octagonal tiling,
 blocks are $not$ all isolated or disjoint as
Figs.2b and c show. Blocks as defined
above can share pairs of sites, with a finite frequency of occurrence.
 This makes the tiling, a true two-dimensional
structure, harder to solve than a fractal Sierpinski-type structure,
which would have less connectivity. 
Fig.3 shows the tiling with grey dashes connecting such pairs
of shared sites. To disconnect the clusters along the grey
lines, the two spins are assigned to one or other of the overlapping blocks.
This is done by
$annulling$ one of the couplings for each of the spins.
Thus we have a diluted version of the original tiling, with certain
couplings annulled. The fraction of annulled couplings is finite,
 and can be calculated exactly to be $\sqrt{2}/\lambda^3 \approx
0.10$ or ten percent.
\begin{figure}[h]
\centerline{\psfig{figure=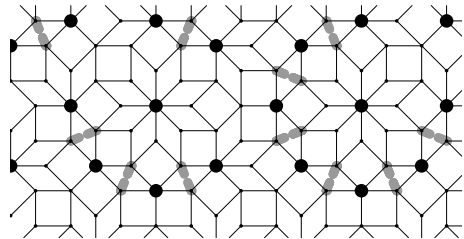,width=6.0cm,angle=0}}
\vspace{.2cm}
\caption{Tiling showing block centers (fat dots). The grey lines 
connect pairs of sites that are shared between two blocks.
}
\end{figure}

For example, Fig.2c shows eight overlapping $D_1$-clusters surrounding an A 
site. They are decoupled by annulling one of the couplings on either side of
each D site. The result is that the coordination numbers of all $D_1$ sites
after dilution goes from $z=5$ to $\tilde{z}=3$. Similarly, C sites
have their $z$ reduced from 6 to 5.

For the seven local environments on the
octagonal tiling, one has specific connectivity rules: an
A site is always coupled to eight F sites, a B site always coupled to
five F sites and 2 E sites, etc. 
The new block spin
variables take on environment dependent values, and after
one inflation,
 one finds that ${\bf{S}}^{(1)} =
(S^{(1)}_A,...,S^{(1)}_F) = C {\bf{S}}^{(0)}$, with
\bea
C = \left(
\begin{array}{rrrrrrr}
-1&0&0&0&0&0&8\\
-1&0&0&0&0&0&8\\
-1&0&0&0&0&0&8\\
-1&0&0&0&0&0&8\\
0&-1&0&0&0&2&5\\
0&0&-1&0&0&3&2\\
0&0&0&-1&1&2&0\\
\end{array}
\right)
\eea
where ${\bf{S}}^{(0)} = (s_0,s_0,....,s_0)$ and $s_0 =\frac{1}{2}$.
One can follow the renormalisations as the tiling undergoes successive
inflations, with ${\bf{S}}^{(n)} = C {\bf{S}}^{(n-1)}$,
and it is easy to show that for large $n$ ${\bf{S}}^{(n)} \approx
3{\bf{S}}^{(n-1)}$, and spins tend to relative asymptotic values given by 
 $(1,1,1,1,1,\frac{3}{4},\frac{1}{2})$.
In a star cluster where all spins have different lengths, 
with $z$ spins of $S_i = n_i s_0$ are coupled by the same 
 $J$ to a central spin $S_0 = 
n_0 s_0$ the ground state energy is (for $\sum_z n_i > n_0$)
\begin{equation}
\epsilon(J,z,\{n\}) = -n_0J(\sum_{i=1}^z n_i + 2)/4 
\end{equation}

If ${\bf{n}}=(n_A,n_B,n_C,n_{D1})$ are the number of
blocks in a given region of each given type, the
number of blocks of each type after one deflation is $N {\bf{n}}$
where
\be
N = \left(
\begin{array}{rrrr}
1&0&0&8 \\
1&0&2&5\\
1&0&4&2\\
1&1&4&0
\end{array}
\right)
\ee
whose largest eigenvalue is equal to 7 so that
the total number of blocks increases(decreases) with the number $m$ of
deflations(inflations) as $7^m$ for large $m$. 

 The effective interaction between block spins is determined by
 inspecting how links transform under inflation. A minimal model can be defined
by considering just five types of links. The set of couplings retained in the
model is represented in an array
${\bf{j}}=(j_{\alpha F}$,$j_{\alpha E}$,$j_{D_1 D_2}$,
$j_{D_2F}$,$j_{EF})$. Here,
$j_{\alpha F}$ is used to denote the link between (A,F), (B,F),(C,F) and
($D_1$,F) pairs. Similarly, $j_{\alpha E}$ denotes the link
connecting (B,E),(C,E) and ($D_1$,E) pairs. After inflation, the
new couplings between sites are written in terms of the
five old couplings, giving rise to a multiplicative
renormalization scheme \cite{note3}.
After one step of
inflation the new couplings (the grey lines in Figs.2) 
are found to be ${\bf{j}}^{(1)} = M^{(0)} {\bf{j}}^{(0)}$ where
\begin{equation}
M^{(n)} = \left(\begin{array}{ccccc}
0&0&0&0&2 \xi^{(n)}_A \xi^{(n)}_D \\
0&0&0&0&3\xi^{(n)}_A\xi^{(n)}_C\\
0&0&0&0&4\xi^{(n)}_A \xi^{(n)}_B \\
0&\xi^{(n)}_B\xi^{(n)}_D&0&\xi^{(n)}_B\xi^{(n)}_D&\xi^{(n)}_B\xi^{(n)}_D\\
0&\xi^{(n)}_C\xi^{(n)}_D&0&\xi^{(n)}_C\xi^{(n)}_D&\xi^{(n)}_C\xi^{(n)}_D
\end{array} \right)
\end{equation}
with the initial condition (taking the zero order
coupling $J=1$) 
 ${\bf{j}}^{(0)}=(1,1,1,1,1)$. 
 After one inflation
the new Hamiltonian is written
using averaged values of the renormalization factors ${\xi^{(1)}_i}$
and couplings since the effective spins and
couplings are no longer uniform. Block energies 
 are obtained using
 Eq.2 with a site-averaged value of the coupling, calculated
appropriately for each of the seven families of sites 
(neglecting minor differences
of local environment in some cases).
A sites, for example, have eight
A-F links to their neighbors, so
their average local coupling is $\overline{\jmath}^{(n)}_{A}  =
 j^{(n)}_{\alpha F}$.
 The matrix $M^{(n)}$ evolves under successive inflations to a fixed point
whose maximum eigenvalue $\gamma_5 \approx 0.15$. Thus for large $n$,
couplings decay as ${\bf{j}}^{(n)}=\gamma_5 {\bf{j}}^{(n-1)}$, while
the corresponding eigenvector 
determines the fixed point relative couplings.

With these definitions, we now turn to the results obtained.
The first quantity of interest is the ground state energy per site, $e_{0}$.
The QMC data in \cite{wess} yield a value of $e_{0} \approx
-0.66$, while that of the square lattice was determined numerically
\cite{carl}
 to be about $-0.67$. A plausible conjecture is that the octagonal tiling, with
its two sublattice structure and its average coordination number of 4 has
the {\it same}
 GS energy as the square lattice, however this remains to be proven.
In the RG scheme, 
the ground state energy can be written as an infinite sum
\bea
e_{0} = \sum_{i \in \alpha} f_i (\epsilon^{(0)}_i + 
\epsilon^{(1)}_i /\lambda^{2}
 ... + \epsilon^{(n)}_i/\lambda^{2n} + .....)
\eea

using the fact that the frequency of blocks of type $i$ is initially
 $f_i$\cite{note2},
 and is diminished by $1/\lambda^{2}$ at each step of RG. Here $\epsilon^{(n)}$ is
 shorthand notation for the energy of an $n$th stage block having
a spin $S_0^{(n)}$ at the center, an averaged coupling value $\overline{\jmath}^{(n)}$,
and surrounding spins of value $S_i^{(n)}$ as given by Eq.2. 
 The series
for the energy gives $e_{0} \approx -0.51$ (compared with the
result of about -0.54 \cite{sierra} for the square lattice).
One reason for the discrepancy between renormalization and
Quantum Monte Carlo results for the quasiperiodic model is 
the appreciable bond dilution occurring at C and D sites,
which leads to having fewer energy terms in the Hamiltonian.
 On the other hand, the loss of
bonds is partly offset by the fact
that the dilution tends also to suppress frustration 
and raise the local order parameter. A crude way to put back the
``missing bond-energies" is to add in $half$ of the missing link energies
at each of the C and D sites. This is easily done by adjusting the
$\tilde{z}$ values ($\tilde{z}_C$ goes up from 5 to 
5.5 while 
$\tilde{z}_{D1}$ is increased from 3 to 4). One then finds an adjusted ground state energy of
about $-0.59$. This correction technique will be applied to the calculation
of local order parameters discussed below, with good results.
 
The  QMC data in \cite{wess} give values of
 local order parameters
defined in terms of the local energies  $E_i = J\sum \langle \vec{S}_i.
\vec{S}_{i+\delta}\rangle$, where the sum is over all nearest neighbors of 
a given site $i$ and the spin correlations are evaluated in the ground state.
The relation taken for the local order parameters is $m_{s,i} = \sqrt{E_i/z}$
\cite{note4}.
We are therefore interested in the cluster energies, $E_i$, 
as a function of $z$. In the zeroth approximation the cluster energies
are just the HS $\epsilon_i^{(0)}$. 
The RG allows us to calculate the cluster energy $E^{(n)}$ on
increasingly bigger length scales, where $z$ is the coordination number
of the central spin at the end of $n$ steps. 
For example, taking $n=1$,
 consider an A-site of the inflated tiling with eight F
sites around it. The ancestor of the central A-site is 
an A site with an associated block energy 
of $\epsilon^{(0)}_{A}$. The ancestors of the
neighbors are $D_1$ sites, and each one contributes half its
block energy to the cluster energy.
 The total energy $E^{(1)}_A$ 
is therefore the sum of the Heisenberg star energy for A sites having a
 first-order coupling $\overline{\jmath}^{(1)}_A$, plus a zeroth order 
$A$ block energy term, plus half the zeroth order
block energies of its neighbors as follows:
\begin{equation}
E^{(1)}_A = \epsilon^{(1)}_A+ \epsilon^{(0)}_A + 4 \epsilon^{(0)}_{D1}
\end{equation}
and similar expressions are written for the other six types of site.
At each stage of RG, the cluster energies $E^{(n)}$ are used to
find the corresponding values of the local order parameters.
For $n=2$, cluster energies for the twice-inflated
tiling can be written out in 
terms of the energies $\epsilon^{(k)}_i$ ($k=0,1,2$). 
The number of terms contributing to the
cluster energy is governed by the largest eigenvalue
of $N$, so that
$E^{(n)}/7^n$  tends to a limit
as $n\rightarrow \infty$.  In that limit,
 the quantities $m_s = \sqrt{z^{-1}E^{(n)}/7^n}$  therefore
have asymptotic values which are compared to
the available numerical data.
 In fig.4a we have compared the $m_{s}$ obtained after zero 
(dashed line), 
one and two and three RG steps (open circles, squares and filled
circles respectively).
After two steps, the values of $m_{s}$ converge quickly.
In Fig.4b are shown the third (circles) and fourth order (squares)
results which overlap on the scale of the figure.
The limiting values of $m_{s}$ are
clearly below the QMC data, shown as grey circles, and this is expected due
to the bond dilution.
If the bond reduction is compensated by putting back half the bonds as we
did earlier to estimate $e_0$, we get estimates for $m_{s}$ values
on the original octagonal tiling. The grey squares of Fig.4b were obtained by
correcting the $n=4$ data in this way, in fairly good agreement
with the QMC data.
\begin{figure}[h]
\centerline{\psfig{figure=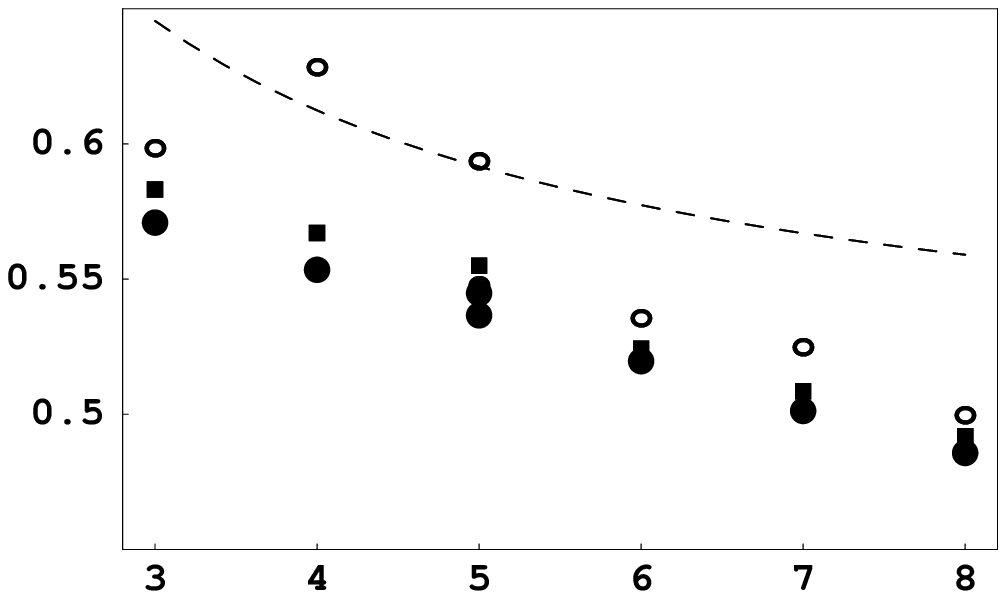,width=6.0cm,angle=0}}
\centerline{\psfig{figure=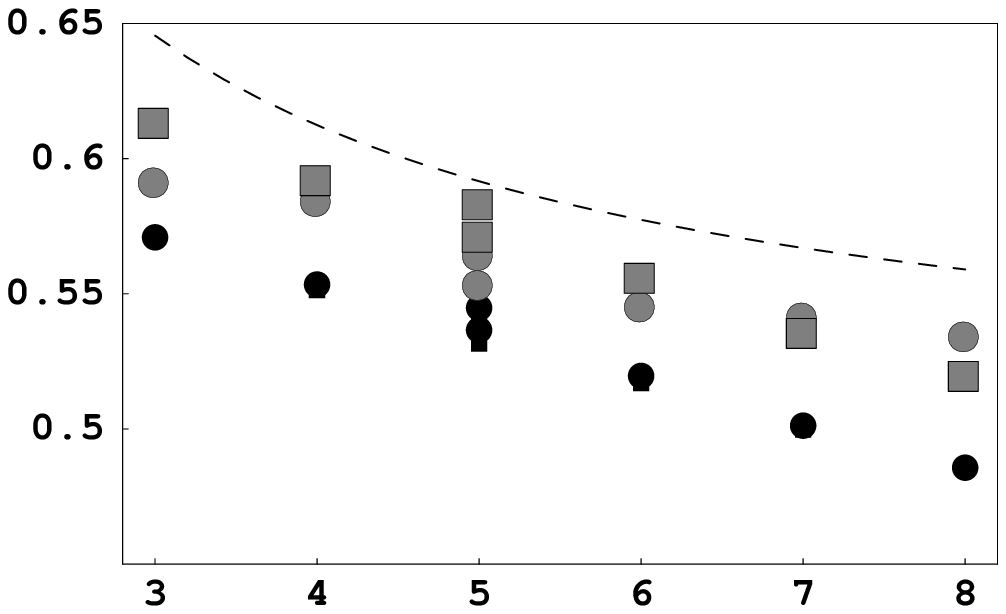,width=6.0cm,angle=0}}
\vspace{.7cm}
\caption{ $m_s$ values plotted versus $z$ for increasing orders of RG.
 (a) zero (dashed line), 1st (open circles), 2nd (rectangles) and
3rd (filled circles) order RG. 
 (b) QMC data (grey circles), 3rd (circles) and 4th (rectangles) order
 RG, and corrected 4th order data (grey rectangles).
}
\end{figure}

In conclusion, we have presented an RG scheme for a two-dimensional
quasiperiodic tiling
that can be completely solved, under certain approximations.
The problem is of importance as 
being the simplest 2d aperiodic quantum antiferromagnetic spin model possible,
like the square lattice
antiferromagnet for periodic systems.
The results of the approximate RG as regards the local order parameters
 are close to those calculated for
the full undiluted model, and we believe the model takes
into account important aspects of
 the quasiperiodic geometry of the tiling. The method is
less good at obtaining the ground state energy. We note that the ground states
 of the
 octagonal tiling
and the square lattice appear to share the same value of the ground state
energy - a conjecture awaiting proof.
One notes finally that the RG calculation
 after appropriate modifications can be extended to
electronic, vibrational and other discrete problems (work in progress).


\begin{references}
\bibitem{note} The two D-types can be distinguished, for example,
 at next-nearest-neighbor level.
\bibitem{note3} Inter-cluster bonds
 are not considered to be affected by bond-dilution.
Thus a central A site linked to eight $D_1$-clusters around it
by sixteen links retains all those links to the eight clusters - i.e.
retains two links per $D_1$-cluster (in a sort of bond-moving step).
\bibitem{note2}
The frequencies of the $\alpha$ class sites are
$f_A=\lambda^{-4}$, $f_B=\lambda^{-5}$, $f_C=2\lambda^{-4}$,
$f_{D1}=\lambda^{-3}$.
\bibitem{note4} Note a slight change of definition of $m_{s}$ 
from that of ref.3.
\end{references}
\end{document}